\documentclass[a4paper]{article}

\usepackage{fullpage}
\usepackage{graphicx}

\usepackage{microtype}

\usepackage{amssymb}
\usepackage{amsmath}
\usepackage{amsthm}

\usepackage{xspace}
\usepackage{color}

\usepackage{todonotes}

\newtheorem{theorem}{Theorem}

\newtheorem{lemma}[theorem]{Lemma}

\let\leq\leqslant
\let\geq\geqslant

\let\theta\vartheta

\newcommand{\Reals}{\mathbb{R}}
\newcommand{\Rd}{\Reals^{d}}
\newcommand{\A}{\mathcal{A}}

\newcommand{\dup}{\lceil d/2\rceil}
\newcommand{\ddown}{\lfloor d/2 \rfloor}
\newcommand{\podim}{\mathrm{dim}}
\newcommand{\hh}{\bar{h}}
\newcommand{\eps}{\varepsilon}

\usepackage{hyperref}
\hypersetup{%
  breaklinks,%
  colorlinks=true,%
  linkcolor=[rgb]{0.45,0.0,0.0},%
  citecolor=[rgb]{0,0,0.45}
}

\makeatletter
\partopsep\z@ \textfloatsep 10pt plus 1pt minus 4pt
\def\section{\@startsection {section}{1}{\z@}%
  {-3.5ex plus -1ex
    minus -.2ex}{2.3ex plus .2ex}{\large\bf}}
\def\subsection{\@startsection{subsection}{2}%
  {\z@}{-3.25ex plus
    -1ex minus -.2ex}{1.5ex plus .2ex}{\normalsize\bf}}
\def\@fnsymbol#1{\ensuremath{\ifcase#1\or 1\or 2\or 3\or 4\or
    5\or 6\or 7 \or 8\ or 9 \or 10\or 11 \else\@ctrerr\fi}}
\makeatother

\title{Better Late than Never: the Complexity of Arrangements of Polyhedra}

\author{Boris Aronov%
  \thanks{Department of Computer Science and Engineering, Tandon
    School of Engineering, New York University, Brooklyn, NY,
    USA. boris.aronov@nyu.edu} 
  \and
  Sang Won Bae%
  \thanks{Division of Computer Science and Engineering, Kyonggi
    University, Suwon, Korea. swbae@kgu.ac.kr}
  \and
  Sergio Cabello%
  \thanks{Faculty of Mathematics and Physics, University of Ljubljana, Slovenia, 
    and Institute of Mathematics, Physics and Mechanics, Slovenia. 
    sergio.cabello@fmf.uni-lj.si}
  \and
  Otfried Cheong%
  \thanks{SCALGO, Aarhus, Denmark. otfried@scalgo.com}
  \and
  David Eppstein%
  \thanks{Computer Science Department, University of California,
    Irvine, USA. eppstein@uci.edu} 
  \and
  Christian Knauer%
  \thanks{Department of Computer Science, University of Bayreuth,
    Germany. christian.knauer@uni-bayreuth.de}
  \and
  Raimund Seidel%
  \thanks{Saarland University, Saarland Informatics Campus, Saarbrücken, Germany.
    rseidel@cs.uni-saarland.de}%
}

\begin{document}

\maketitle

\begin{abstract}
  Let~$\A$ be the subdivision of~$\Rd$ induced by $m$~convex polyhedra
  having~$n$ facets in total.  We prove that~$\A$ has combinatorial
  complexity~$O(m^{\dup} n^{\ddown})$ and that this bound is tight.
  The bound is mentioned several times in the literature, but no proof
  for arbitrary dimension has been published before.
\end{abstract}

\section{Introduction}

We consider a collection of~$m$ convex polyhedra in~$\Rd$, each given
as the intersection of halfspaces, where the total number of
facets is~$n \geq m$ and where we consider the dimension~$d$ to be
a constant.  The family of polyhedra induces a subdivision~$\A$
of~$\Rd$ into cells and faces of dimensions~$0$ to~$d$.  What is the
complexity of this subdivision~$\A$, that is,~what is the number of
its faces?

When $m = 1$, $\A$ is a single convex polyhedron defined by~$n$
halfspaces, so by the Upper Bound Theorem~\cite{mcmullen1970}, its
complexity is~$O(n^{\ddown})$.  At the other extreme, when~$m = n$,
each polyhedron is a halfspace, and~$\A$ is the arrangement of~$n$
hyperplanes, which has complexity~$\Theta(n^{d})$.

We generalize both bounds by showing
\begin{theorem}
  \label{thm:main}
  The subdivision in~$\Rd$ induced by~$m$ convex polyhedra with a
  total of~$n$ facets, has complexity~$O(m^{\dup} n^{\lfloor d/2
    \rfloor})$, and this bound is tight.
\end{theorem}

This bound has been mentioned several times in the
literature~\cite{aronov1997common, aronov1997union, berg1997sparse,
  guibas1998polyhedral, guibas1997robot}, referring to an unpublished
manuscript by Aronov, Bern, and Eppstein from some time between~1991
to~1995.  The authors no longer have a copy of the original
manuscript, and do not recollect their proof.  The second
edition~\cite{toth2004handbook} of the \emph{Handbook of Discrete and
Computational Geometry} describes the bound in Section~24.6.  The
chapters on arrangements by Agarwal and
Sharir~\cite[page~58]{agarwal2000arrangements} and Pach and
Sharir~\cite[page~19]{pach2009combinatorial} also state the bound.

Hirata et al.~\cite{Hirata} claim a slightly weaker result in their
Lemma~4.2, but the last line of their proof does not hold in dimension
larger than four.

After we presented our result in Dagstuhl, Dan Halperin found a copy
of the original manuscript in his personal archive.  The first three
pages of this manuscript are reproduced here in the appendix: the
lower-bound construction is identical to ours, while their upper-bound
proof is actually much simpler, using the Upper Bound Theorem as a
black box, while our new proof generalizes the proof of an asymptotic
version of the Upper Bound Theorem by Seidel~\cite{Seidel1995}.  The
original manuscript contains no stringent treatment of degenerate
cases, in fact, their statement ``perturbing the input will increase
the complexity'' is false in this simple form, see
Section~\ref{sec:upper2}.

\section{The upper bound, assuming general position}
\label{sec:upper}

We first assume that the hyperplanes defining the~$m$ polyhedra are in
general position: no two facets lie in the same hyperplane, and the
arrangement of the $n$~hyperplanes is simple, that is, every~$d$
hyperplanes intersect in a single point.  Since every face of the
subdivision~$\A$ has at least one vertex, and since every vertex is
incident to a constant number of faces, this implies that the number
of faces of~$\A$ is bounded by the same term.  In the next section we
will then generalize this to arbitrary polyhedra by a perturbation
argument.

We pick a generic vertical direction; in particular, no two vertices
of the arrangement will be at equal altitude.

Let~$v$ be a vertex of the subdivision~$\A$.  It is defined by~$d$
hyperplanes~$H$.  Let~$I$ be the set of polyhedra that contribute
to~$H$, and let~$U$ be a neighborhood of~$v$ that is small enough such
that only the hyperplanes of~$H$ intersect it.

The $d$~hyperplanes~$H$ cut~$U$ into~$2^{d}$ cells.  One of these cells
lies in the intersection~$P$ of the polytopes in~$I$.  This
polytope~$P$ has~$d$ edges incident to~$v$. As in
Seidel~\cite{Seidel1995}, we observe that at least half of these edges
either go up or down with respect to our vertical direction. Let us
assume there are $i \geq d/2$ edges going up (vertices where the
majority of edges go down are counted analogously).  Then there is a
unique $i$-face~$f$ in~$P$ that contains those
edges~\cite{Seidel1995}.

The $i$-face~$f$ lies in an $i$-flat~$F$ that is the intersection
of~$d - i$ of the hyperplanes in~$H$.  Let~$H'$ be this subset of
hyperplanes.  The intersection of~$F$ with~$P$ is exactly
the~$i$-face~$f$, and $v$~is the lowest vertex of~$f$.

This implies that the vertex~$v$ is uniquely defined by the choice of
the $d-i$ hyperplanes~$H'$ and the intersection polytope~$P$. The
polytope~$P$, on the other hand, is uniquely defined by the at
most~$d$ polytopes in~$I$.  That is, we can uniquely specify~$v$ by
selecting the~$d-i$ hyperplanes~$H'$ and the at most~$i$ polytopes
that appear in~$H \setminus H'$.

For a given~$i \geq d/2$, there are therefore~$O(n^{d-i} m^{i})$ such
vertices, for a total of
\[
\sum_{i=\lceil d/2 \rceil}^{d} O(n^{d-i} m^{i})
= O(m^{\dup} n^{\ddown}).
\]

\section{The upper bound in the general setting}
\label{sec:upper2}

We now turn to the fully general case, where many facets might
intersect in a single vertex, where an $i$-face can lie in more
than~$d-i$ facets, where facets of distinct polyhedra can lie on the
same hyperplane, and where the polyhedra can be lower-dimensional.

Since lower-dimensional polyhedra have no ``facets'' (in the sense of
$(d-1)$-dimensional faces), we prefer to think about our geometry as a
set~$H$ of $n$~colored halfspaces, where the number of colors is~$m$.
Each polyhedron is the common intersection of the halfspaces of one
color.  Note that the bounding hyperplanes of several colored
halfspaces could be identical.  Let~$\hh$ denote the hyperplane
bounding a halfspace~$h$.

We will convert the subdivision~$\A$ formed by the polyhedra into a
subdivision~$\A'$ where the hyperplanes supporting the facets are in
general position, and show that the number of faces of~$\A'$ is at
least the number of faces of~$\A$.  The result of the previous section
then implies the upper bound in Theorem~\ref{thm:main}.

We start by adding a large simplex~$\Delta$ (that is, $d$~halfspaces
of a new color) to the scene, which contains all the vertices and
intersects all the faces of the subdivision.  In the following, it
therefore suffices to consider the faces of~$\A$ that lie
inside~$\Delta$.

\medskip

Consider an $i$-face~$f$ of a subdivision~$\A$ induced by a family of
colored halfspaces.  The affine hull of~$f$ is an $i$-flat~$F$, which
is the common intersection of \emph{at least} $d-i$~bounding
hyperplanes. Let $H_f$ be the set of hyperplanes that bound~$f$, but
do not contain~$f$.  For each~$\hh \in H_f$, exactly one of the two
closed halfspaces bounded by~$\hh$ contains~$f$.  We denote this
by~$h_f$, and let $P_f = \bigcap_{\hh \in H_f} h_f$.

We observe that $f = F \cap P_f$.  We observe further that the
polyhedron~$P_f$ is full-dimensional.  Indeed, consider a point~$p$ in
the relative interior of~$f$. For each $\hh \in H_f$, $p$ lies in the
interior of~$h_f$ (if it lies in~$\hh$, then $\hh$ contains~$f$, a
contradiction). It follows that a neighborhood of~$p$ lies in~$P_f$,
so~$P_f$ is a $d$-dimensional polyhedron.

\medskip

We will perturb the $n$~colored halfspaces one by one, in arbitrary
order.  At each step, the set of halfspace that have already been
perturbed will be in general position.  When several colored
halfspaces are bounded by a common hyperplane, then each one will be
perturbed separately, so we will end up with a situation where all
$n$~perturbed colored halfspaces have distinct hyperplanes (and the
$n$~hyperplanes will be in general position).

We start with the $d$~colored halfspaces of~$\Delta$. They are in
general position by construction.

Consider now the situation at some step of the process: $H$ is the
current set of colored halfspaces, $\A$ is the subdivision defined
by~$H$, and~$h$ is one of the halfspaces that has not yet been
perturbed.  We will perturb~$h$ into a new halfspace~$h'$, such that
$H' = H \setminus \{h\} \cup \{h'\}$ and~$\A'$ is the subdivision
defined by~$H'$.  We will ensure that the number of faces of~$\A'$
inside~$\Delta$ is at least the number of faces of~$\A$
inside~$\Delta$.

The perturbation ``moves''~$h$ slightly ``outwards.''  Formally, we
pick the halfspace~$h'$ such that (for some~$\eps > 0$ to be
determined):
\begin{enumerate}
\item $h' \cap \Delta \supset h \cap \Delta$, and
\item the distance between~$\hh' \cap \Delta$ and~$\hh \cap \Delta$ is
  at most~$\eps$, and
\item the coefficients of~$\hh'$ are algebraically independent of any
  coefficients of the hyperplanes in~$H$.
\end{enumerate}

We show now that for all faces of~$\A$ inside~$\Delta$ there is a
corresponding face of the same dimension of~$\A'$. Let~$f$ be an
$i$-face of~$\A$ that lies inside~$\Delta$, for $0 \leq i \leq d$.
Note that we consider faces as closed sets.

If~$f \cap \hh = \emptyset$, then~$f$ has a positive distance~$\delta$
from~$\hh$.  By choosing $\eps < \delta$, we ensure that~$f$ remains
unchanged by the perturbation.

If~$f \cap \hh \neq \emptyset$, but~$f$ is not contained in~$\hh$,
then we write $f = F \cap P_f$, for an $i$-flat~$F$ and a
full-dimensional polyhedron~$P_f$.  By assumption, $\hh$ does not
contain~$F$.  If $\hh \not\in H_f$, then~$h$ does not contribute
to~$P_f$, and the perturbation leaves~$f$ unchanged.  Otherwise, $\hh
\in H_f$, so the halfspace~$h_f$ bounded by~$\hh$ and containing~$f$
contributes to~$P_f$.  Pick a point~$p$ in the relative interior
of~$f$.
Since~$P_f$ is full-dimensional, $p$ lies in the interior of~$P_f$,
and so~$p$ has a positive distance~$\delta$ to~$\hh$.  We again
choose~$\eps < \delta$, and let $h'_f$ to be the halfspace bounded
by~$\hh'$ containing~$p$.   Replacing~$h_f$ by~$h'_f$, we obtain a new
polytope~$P'_f$, which again contains~$p$ in its interior, and so~$F
\cap P'_f$ is an $i$-face of~$\A'$.

It remains to consider the case where~$f$ lies in~$\hh$.  Let~$\A_h$
be the subdivision obtained by deleting~$h$ from~$\A$.  There are two
subcases:

If~$f$ exists unchanged in~$\A_h$, then it also exists in~$\A'$.  Here
we need the first perturbation property, which implies that~$f$ lies
entirely in the interior of~$h'$.

Finally, we consider the case where~$f$ does not exist in~$\A_h$.
This implies that there is an $(i+1)$-face~$g$ of~$\A_h$ that
contains~$f$.  We write $g = G \cap P_g$ as above, with~$G$ an
$(i+1)$-flat and~$P_g$ a full-dimensional polyhedron.

By assumption~$f = g \cap \hh = G \cap \hh \cap P_g$.  Since~$P_g$ is
full-dimensional, we can choose~$\eps$ small enough such that~$P_g
\cap \hh' \neq \emptyset$.  But then $G \cap \hh' \cap P_g$ is
an~$i$-face of~$\A'$.

We observe that for distinct faces of~$\A$, the corresponding faces
of~$\A'$ are also distinct.

\medskip

Repeating this argument for each of the~$n$ colored halfspaces, we
obtain a new arrangement~$\A'$ of~$n+d$ halfspace colored with~$m+1$
colors (with one color class forming the simplex~$\Delta)$.

By the third perturbation property, the~$n+d$ hyperplanes are in fully
general position: every~$d$-tuple intersects in exactly one point.
By the argument in the previous section, the arrangement~$\A'$
has~$O(m^{\dup} n^{\ddown})$ faces.  And this implies that the number of faces
of~$\A$ is bounded by the same term as well, proving the upper bound
in Theorem~\ref{thm:main}.

\section{The lower bound}

Recall the product construction for convex polytopes, as for instance described
in Ziegler's book~\cite[page~10]{Ziegler2012}.  For polytopes
$P\subset \Reals^p$ and $Q\subset \Reals^q$ 
the product polytope is defined to be the set
\[
P\times Q = \left\{ \left( {x \atop y} \right) \,:\, x\in P,~y\in Q \right\}\,.
\]
This product polytope has dimension $\podim(P)+\podim(Q)$ and its
nonempty faces are the products of nonempty faces of $P$ and nonempty
faces of $Q$. The inequalities describing the facets of $P\times Q$
are the union of the inequalities describing the facets of $P$ (which
have coefficients 0 for the ``$y$-coordinates'') and the inequalities
for the facets of $Q$ (which have coefficient 0 for the
``$x$-coordinates'').  The coordinates of the vertices of $P\times Q$
are all concatenations of the (``$x$'') coordinates of the vertices
of~$P$ with (``$y$'') coordinates of vertices of $Q$. This implies
that the number of facets of $P\times Q$ is the sum of the facet
numbers of~$P$ and~$Q$, whereas the number of vertices is the product
of the vertex numbers.

It is easy to see that something analogous holds for facet and vertex
numbers of product subdivisions:

\begin{lemma}
 \label{lemma:tool}
 Let $P_1,\ldots,P_m$ be polytopes in~$\Reals^{p}$ with $N$~facets in
 total and with $V$~vertices in the induced subdivision.  Similarly
 let $Q_1,\ldots,Q_m$ be polytopes in~$\Reals^{q}$ with $M$~facets in
 total and with $W$~vertices in the induced subdivision.
 
 The set $P_1\times Q_1, P_2 \times Q_2, \ldots, P_m\times Q_m$ of $(p+q)$-dimensional
 polytopes has $N+M$ facets in total and their induced subdivision has $V\cdot W$ vertices. 
 \end{lemma}

Let $s>1$ be an integer constant. For integer $\ell\geq 3$ let $C$ be a regular $\ell$-sided
convex polygon in $\Reals^2$ with edges tangent to the unit circle.  Consider the $s$-fold
product polytope $C^s = \underbrace{C \times C \times \cdots \times C}_{s~ \mathrm{times}}$.
It has $s\cdot \ell$ facets and $\ell^s$ vertices.  For $n=s\ell$ and $d=2s$ this is a particularly
simple construction of a $d$-polytope with $n$ facets and an asymptotically maximal 
$O(n^{\ddown})$ vertices.

For integer $m\geq 1$ and $0\leq i < m$ let $C_i$ be the polygon $C$ rotated by
$i\frac{2\pi}{\ell m}$ around the origin and let $P_i$ be the $d$-polytope $C_i^s$, where
we continue to consider even $d=2s$.

We claim that the polytopes $P_0,\ldots,P_{m-1}$ have $sm\ell$ facets
in total, and their subdivision has $(\ell \cdot m^2)^s$ vertices.

It suffices to show that the the polygons $C_0,\ldots,C_{m-1}$ have in total
$m\ell$ facets and their induced subdivision has $\ell\cdot m^2$ vertices, and then
repeatedly apply Lemma~\ref{lemma:tool}.
The total facet number for the $m$ polygons is clearly $m\ell$.  
For the vertex count in the subdivision observe that each of the $m\choose 2$ pairs of the
$\ell$-gons have their boundaries intersect in $2\ell$ points, which, including the $\ell m$
 corners yields overall $\ell \cdot m^2$ vertices.

If you let $\ell=\frac{n}{s\cdot m}$, then $P_0,\ldots,P_{m-1}$ have $n$ facets overall, and
the subdivision has
\[
(\ell \cdot m^2)^s = \frac{n^s m^s}{s^s} = \Theta(n^{\ddown}m^{\dup})
\]
vertices if $s$ is considered a constant and we are considering even dimension $d=2s$.

For odd $d=2s+1$, take the above construction, choose $m$ intervals, say, $J_i = [-1-i,1+i]$,
choose $\ell = \frac{n-2m}{ms}$ for the construction of the $P_i$'s above,
and consider the products $Q_i=P_i\times J_i$ for $0\leq i < m$.  The $J_i$ have $2m$
``facets'' in total and their subdivision has $2m$ vertices.  Applying Lemma~\ref{lemma:tool}
then yields that the $Q_i$'s have $n$ facets in total and the number of vertices
in their induced subdivision is
\[
2m \cdot \Theta(n^s m^s) = \Theta(n^s m^{s+1}) = \Theta(n^{\ddown}m^{\dup})\,.
\]

\section*{Acknowledgments}

Work by Boris Aronov was partially supported by NSF Grant
CCF-20-08551; part of the research was done while visiting University
of Bayreuth in the summer of 2024.  S.W. Bae was supported by the
National Research Foundation of Korea (NRF) grant funded by the Korea
government(MSIT) (No.~RS-2023-00251168).  Research of David Eppstein
was supported in part by NSF grant~CCF-2212129.  Sergio Cabello was
funded in part by the Slovenian Research and Innovation Agency
(P1-0297, N1-0218, N1-0285). Research of Sergio Cabello and Otfried
Cheong was funded in part by the European Union (ERC, KARST, project
number~101071836). Views and opinions expressed are however those of
the authors only and do not necessarily reflect those of the European
Union or the European Research Council. Neither the European Union nor
the granting authority can be held responsible for them.

\bibliography{references}

\begin{thebibliography}{10}

\bibitem{agarwal2000arrangements}
Pankaj~K. Agarwal and Micha Sharir.
\newblock Arrangements and their applications.
\newblock In {\em Handbook of computational geometry}, pages 49--119. Elsevier,
  2000.

\bibitem{aronov1997common}
Boris Aronov and Micha Sharir.
\newblock The common exterior of convex polygons in the plane.
\newblock {\em Computational Geometry}, 8:139--149, 1997.

\bibitem{aronov1997union}
Boris Aronov, Micha Sharir, and Boaz Tagansky.
\newblock The union of convex polyhedra in three dimensions.
\newblock {\em SIAM Journal on Computing}, 26:1670--1688, 1997.

\bibitem{berg1997sparse}
Mark~de Berg, Dan Halperin, Mark Overmars, and Marc~van Kreveld.
\newblock Sparse arrangements and the number of views of polyhedral scenes.
\newblock {\em International Journal of Computational Geometry \&
  Applications}, 7:175--195, 1997.

\bibitem{guibas1998polyhedral}
Leonidas~J. Guibas, Dan Halperin, Hirohisa Hirukawa, Jean-Claude Latombe, and
  Randall~H. Wilson.
\newblock Polyhedral assembly partitioning using maximally covered cells in
  arrangements of convex polytopes.
\newblock {\em International Journal of Computational Geometry \&
  Applications}, 8:179--199, 1998.

\bibitem{guibas1997robot}
Leonidas~J. Guibas, Rajeev Motwani, and Prabhakar Raghavan.
\newblock The robot localization problem.
\newblock {\em SIAM Journal on Computing}, 26:1120--1138, 1997.

\bibitem{Hirata}
Tomio Hirata, Jiří Matoušek, Xue-Hou Tan, and Takeshi Tokuyama.
\newblock Complexity of projected images of convex subdivisions.
\newblock {\em Computational Geometry}, 4:293--308, 1994.
\newblock \href {https://doi.org/10.1016/0925-7721(94)00009-3}
  {\path{doi:10.1016/0925-7721(94)00009-3}}.

\bibitem{mcmullen1970}
Peter McMullen.
\newblock The maximum numbers of faces of a convex polytope.
\newblock {\em Mathematika}, 17(2):179--184, 1970.

\bibitem{pach2009combinatorial}
J{\'a}nos Pach and Micha Sharir.
\newblock {\em Combinatorial geometry and its algorithmic applications: The
  Alcal{\'a} lectures}.
\newblock American Mathematical Soc., 2009.

\bibitem{Seidel1995}
Raimund Seidel.
\newblock The upper bound theorem for polytopes: an easy proof of its
  asymptotic version.
\newblock {\em Computational Geometry}, 5:115--116, 1995.
\newblock \href {https://doi.org/10.1016/0925-7721(95)00013-Y}
  {\path{doi:10.1016/0925-7721(95)00013-Y}}.

\bibitem{toth2004handbook}
C.D. Toth, J.~O'Rourke, and J.E. Goodman.
\newblock {\em Handbook of Discrete and Computational Geometry, Second
  Edition}.
\newblock Discrete Mathematics and Its Applications. CRC Press, 2004.
\newblock URL: \url{https://books.google.de/books?id=X1gBshCclnsC}.

\bibitem{Ziegler2012}
G\"unter~M. Ziegler.
\newblock {\em Lectures on Polytopes}.
\newblock Springer New York, NY, 2012.

\end{thebibliography}

\appendix

\section{The original manuscript}

\includegraphics[page=1,trim=20mm 35mm 20mm 40mm,clip]{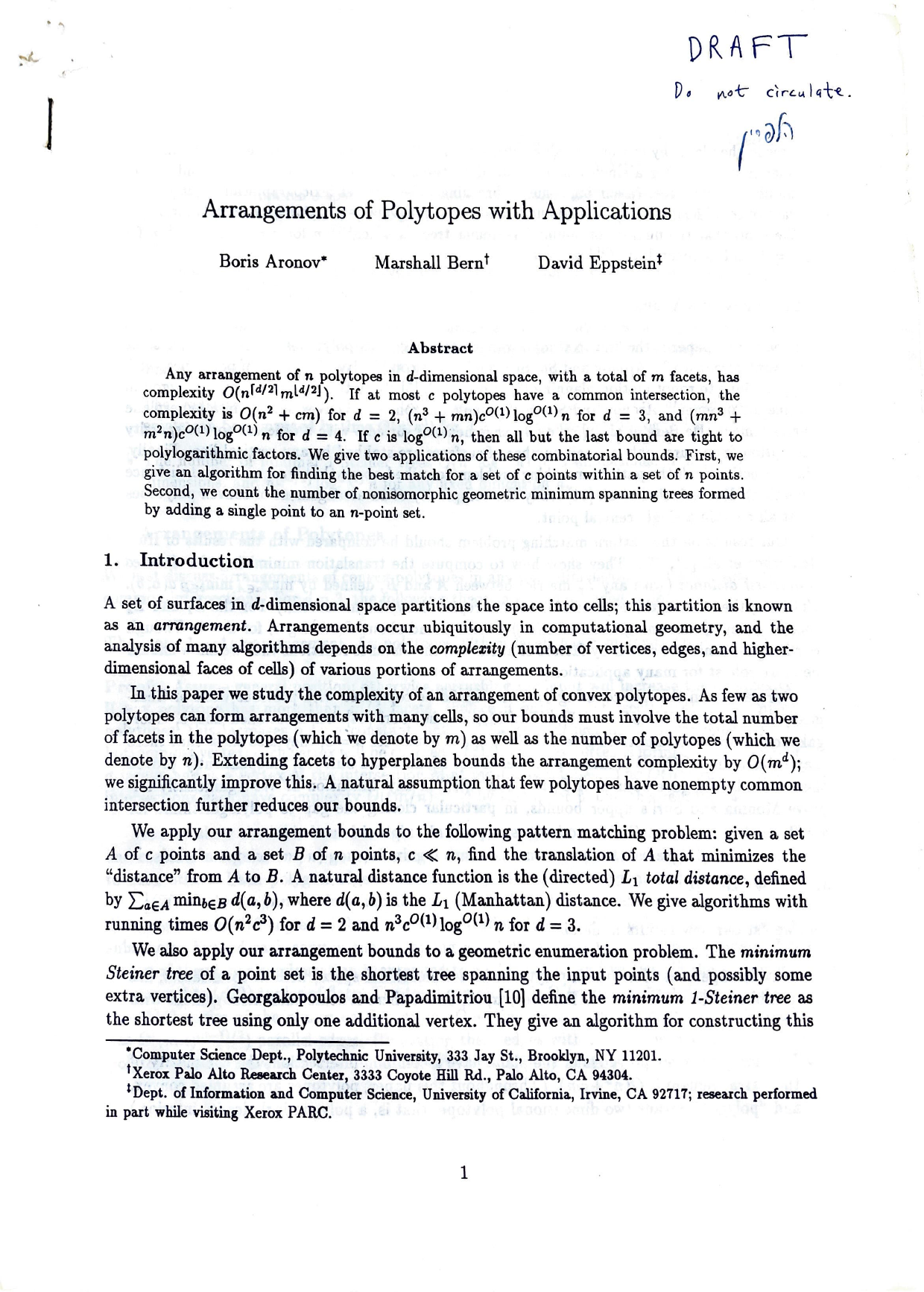}

\includegraphics[page=2,trim=23mm 35mm 20mm 30mm,clip]{historic.pdf}

\includegraphics[page=3,trim=10mm 35mm 20mm 30mm,clip]{historic.pdf}

\end{document}